\documentclass[11pt]{article}

\usepackage{
amssymb,epsf,amsmath}

\numberwithin{equation}{section}
\def\Ann {{\rm{Ann}}}

\begin{document}
\setcounter{page}{0}
\begin{titlepage}
\titlepage
\rightline{hep-th/0512185}
\rightline{SPhT-T05/202}
\rightline{CPHT-RR  073.1205}
\vskip 3cm

\centerline{\Large{Tachyon condensation and D-branes in generalized geometries}}
\vskip 1.3cm
\centerline{Pascal Grange $^a$ and Ruben Minasian $^{b,c}$}

\begin{center}
\em$^a$  School of Natural Sciences, Institute for Advanced Study\\
Princeton, NJ 08540,  USA\\
\vskip .4cm
$^b$ Service de Physique Th{\'e}orique, CEA/Saclay \\
 91191 Gif-sur-Yvette, France\\
\vskip .4cm
$^c$ Centre de Physique Th{\'e}orique, \'Ecole
Polytechnique
\\91128 Palaiseau Cedex, France\\
\end{center}
\vskip 1.5cm

\begin{abstract}

In generalized complex geometry, D-branes can be seen as maximally
isotropic spaces and are thus in one-to-one correspondence with pure
spinors.  When considered on the sum of the tangent and cotangent
bundles to the ambient space, all the branes are of the same dimension
and the transverse scalars enter on par with the gauge fields; the
split between the longitudinal and transverse directions is done in
accordance with the type of the pure spinor corresponding to the given
D-brane. We elaborate on the relation of this picture to the
T-duality transformations and stability of D-branes.  A discussion of
tachyon condensation in the context of the generalized complex
geometry is given, linking the description of D-branes as generalized
complex submanifolds to their K-theoretic classification.

\end{abstract}

\vfill
\begin{flushleft}
{\today}\\
\end{flushleft}
\end{titlepage}


\newpage


\section{Introduction}

In three years since its appearance, the generalized complex geometry
(GCG) \cite{hitchin,gualtieri} has had a variety of applications of
string theory. The formalism is distinguished by a number of features,
such as the parity between $B$-field and diffeomorphisms, and the
resulting interpolation between complex and symplectic geometries;
there is also a natural twisting by a three-form $H$-flux. All this
makes GCG a very natural framework for addressing questions like
classification of supersymmetric flux backgrounds or extending the
notion of mirror symmetry beyond Calabi--Yau manifolds.\\

The question of the proper definition of D-branes in the context of
 generalized complex geometry has also received some attention, and
 some proposals and results have appeared, based on inner geometric
 consistency and on physical realizations of generalized complex
 geometry through localization in supersymmetric sigma-models
 \cite{lmtz,zabzine}.  Generalized submanifolds have indeed been
 proposed as Abelian generalized complex D-branes by Gualtieri with
 the motivation of covariance with respect to the $B$-field
 transformation. Furthermore, studies of (2,2) theories with
 boundaries have shown the relevance of these objects from the point
 of view of sigma-models.  Different aspects of D-branes in GCG have
 been discussed in \cite{BBB,NCkapustin,mezigue,cgj,zucchini,openbrst}. \\

More relevantly for the present discussion, the development of
generalized complex geometry has led to some unification between the A
and B models of topological strings \cite{NCkapustin,sigma}. Since as mentioned the
framework naturally incorporates the $B$-field, the inclusion of
D-branes can be inspired by the gauge-equivalent picture of a field
strength. In particular, it was shown \cite{gm} that stable B-branes
\cite{MMMS} are mapped by mirror symmetry to stable A-branes,
including those supported on non-Lagrangian submanifolds. Such an
incorporation of Fourier--Mukai transform into the mapping between
pure spinors \cite{fmt} (see also \cite{BB, ale}) illustrates the unifying power of the
generalized geometry.\\

One characteristic feature of describing the D-branes as generalized
submanifolds is that the gauge fields and the transverse scalars enter
on the same footing (just as forms and vectors, or $B$-field and
diffeomorphisms) and there is no distinction between big (higher-dimensional)
and small (lower-dimensional) branes. Thus the understanding of the D-branes in
generalized geometries should eventually lead to natural incorporation
of phenomena typically described via T-duality (see 
e.g. \cite{taylor}).  These will be in the focus of the attention of the present paper.\\

 As such the discussion of D-branes here is decoupled from issues
related to preservation of supersymmetry; yet the integrability of a
given generalized structure is intimately connected with supersymmetry
of flux backgrounds \cite{n=1}. See \cite{koerber, MS} for a related
discussion of supersymmetric generalized D-branes.  Generalized
submanifolds are locally graphs over gauge bundles. It is interesting
to inquire about the consistency and possible relations of this
description of D-branes with e.g. K-theoretic classification of
D-brane charges. As a first step in this direction, we will
concentrate here on the transformation properties involving a change
of dimension, either by duality or by some dynamical process. So after
a discussion of T-duality transformations of generalized complex
branes in section 2, we will address in section 3 the question how
they can be subject to tachyon condensation \cite{sen}.  We will have to face the
lack of a definition for tachyon fields in this framework. We expect
that compatibility conditions such as the Hermitian Yang--Mills
equations can be generalized in a way that the symmetry between the
ordinary gauge bundles and the one-form coordinates is restored,
corresponding to the symmetry of the winding states and momenta on the
D-brane worldvolume. The two questions of T-duality and tachyon
condensation are tied together in discussing the mirror symmetry of
stable triples in section 4, where the role of winding numbers in
generalized geometries is also illustrated.

\section{Generalized complex submanifolds, T-duality and D-branes}

In this section, we will collect a few basic ingredients for
describing D-branes in generalized complex geometry. The proposal by
Gualtieri \cite{gualtieri} for branes as generalized submanifolds will
be our starting point.\\

Let us first review the generalized linear algebra. Consider an
$n$-dimensional vector space $V$ and its dual $V^\ast$. Going from
linear algebra to differential geometry, these will eventually be the
local tangent and cotangent spaces of an $n$-dimensional manifold
$M$. The pairing defined on $V\oplus V^\ast$ by 
\begin{equation}\label{pairing}\langle X+\xi, Y+\eta \rangle=\frac{1}{2}(\iota_X\eta+\iota_Y\xi),\end{equation} 
with $X,Y\in V$ and $\eta, \xi\in V^\ast$, has signature $(n,n)$. A
  generalized almost complex structure on $V\oplus V^\ast$ is an
  almost complex structure that is orthogonal with respect to this
  pairing.  A null space with respect to this pairing, or isotropic
  subspace of $V\oplus V^\ast$, has dimension $n$ at most. An
  $n$-dimensional isotropic subspace is therefore called maximally
  isotropic.\\

 Examples of maximally isotropic subspaces are given by graphs over a
 subspace $E$ of $V$ in the following form:
\begin{equation}L(E,F):=\{X+\xi\in E\oplus V^\ast,\, \xi|_E=\iota_X F\},\end{equation}
where $F$ is a two-form in $\Lambda^2 E$, or equivalently a map from $E$ to $E^\ast$. Indeed by virtue of
(\ref{pairing}), such a graph is isotropic
\begin{equation}\langle X+\iota_X F, Y+\iota_Y F\rangle  =\frac{1}{2}(\iota_X\iota_Y+\iota_Y\iota_X) F=0,\end{equation}
and it is maximally isotropic since it is defined by $\mathrm{dim}\,E$
equations in $V^\ast$ over the basis $E$. Moreover, it was shown in
\cite{gualtieri} that every maximally isotropic space is a graph of
this form.\\

 The $B$-field transformation acts on $V\oplus V^\ast$ (without
 transforming the vector part), as
 \begin{equation} X+\xi\mapsto X+\xi+\iota_X B.\end{equation}
This reads as the action of the exponential $e^B$, which in a
basis adapted to the sum $V\oplus V^\ast$, takes the form
\begin{equation} e^B=
\begin{pmatrix}
1 & 0 \\
B & 1 \hfill 
\end{pmatrix}\end{equation}
The $B$-field transformation induces a transformation of
maximally isotropic subspaces that preserves the projection on $V$,
 as
\begin{equation}\label{covariance} e^B L(E,F)=L(E, B+F),
\end{equation}
in which sense maximally isotropic subspaces are covariant w.r.t. the
$B$-field transformation. The codimension of $E$ is called the type of
the maximally isotropic subspaces $L(E,F)$, and it is invariant under
both diffeomorphisms and $B$-field transformations. The subspace
$L(E,0)$ is the direct sum of $E$ and its annihilator in $V^\ast$:
\begin{equation} L(E,0)=E\oplus \Ann E. \end{equation}

 The exterior algebra $\Lambda^\bullet V^\ast$ carries a representation of
 ${\mathrm{Clifford}}(n,n)$ through the action of $V\oplus V^\ast$ on
 a sum $\phi$ of differential forms:
 \begin{equation} (X+\xi).\phi=\iota_X \phi+ \xi\wedge\phi,\end{equation}
 \begin{equation} (X+\xi).((X+\xi).\phi)= \langle X+\xi, X+\xi\rangle
\phi.\end{equation} Given a spinor, that is, a sum of differential
 forms, one can associate to it its null space in $V\oplus
 V^\ast$. Maximally isotropic subspaces are therefore in one-to-one
 correspondence with pure spinors \cite{chevalley}. Moreover, pure spinors can be
 represented in a way\footnote{This is seen by first noting that
 $L(E,0)$ is the null space of $\det(\Ann E)$, and then by
 $B$-transforming both objects, with $B=F$.} that makes the
 correspondence manifest:
 \begin{equation}\label{forms} L(E,F)\sim \det(\Ann E)\wedge e^F,\end{equation}
where we have used the symbol $\sim$ to indicate that the {\sc{RHS}}
 is a representative of a pure spinor {\emph{line}}, that is defined
 up to a multiplicative factor.  The differential form $\det(\Ann E)$
 is the wedge product of one-forms of any basis of $\Ann E$; as
 mentioned its dimension (type) is an invariant of the maximal
 isotropics.\\

 Given this geometric set-up, there is a natural definition for
{\emph{generalized complex branes}} (GC branes). These objects must be
supported on submanifolds of a GC manifold, subject to a compatibility
condition with the ambient generalized complex structure
$\mathcal{J}$. Extending from the linear algebra above to the sum of
the tangent and cotangent bundles of an $n$-dimensional manifold $M$,
let the generalized tangent bundle of a submanifold of $M$, still
denoted by $E$, carrying a $U(1)$-bundle with gauge curvature $F$, be
\begin{equation}\label{gtbundle}\tau_E^F:=\{X+\xi\in TE\oplus T^\ast M|_E,\, \xi=\iota_X F\}.\end{equation}
This is a maximally isotropic subspace, the $B$-transform (with $B=F$)
of $\tau_E^0$. Demanding that it is
stable under the action of the GC structure makes it the tangent
bundle of a GC brane. The covariance property (\ref{covariance})
w.r.t. the $B$-field transformation is what motivated the definition
of generalized complex branes in \cite{gualtieri}. As in the linear
case, it can be shown that every maximally isotropic subspace is of
this form.\\

This definition is in accordance with the more conventional definitions of D-branes.
Indeed, the two cases where $\mathcal{J}$ is diagonal or anti-diagonal in a
 basis adapted to the direct sum of tangent and cotangent spaces,
 corresponding to the cases where $\mathcal{J}$ comes from a complex
 structure (on $M$) and from a symplectic structure (on $M$) respectively, were worked out in chapter
 7 of \cite{gualtieri}. The generalized complex submanifolds yield respectively
 the holomorphic  bundles corresponding to B-branes, and A-branes of all possible
 forms, either of Lagrangian or non-Lagrangian, that had been
 previously derived by world-sheet techniques
 \cite{kapustin--orlov}.\\

 Whereas $B$-transformations cannot change the type of a pure spinor,
T-duality should, since it acts on pure spinors so as to give the
usual changes of dimension on the corresponding branes.  Having
identified $E$ with a D-brane worldvolume, we can see that the type
(i.e. $\dim(\Ann E)$) gives the dimension of the transverse space.
This is more than a numerical coincidence and shortly we will
elaborate the relation of $\det(\Ann E$) with the characteristic
polynomial of transverse displacements.  \\

\subsection{T-duality and generalized complex branes: the D6-brane case}
Let us specialize to the case $n=6$ , and consider the Chern character
encoding a space-filling D6-brane together with lower brane charges \cite{within}
smeared along its worldvolume $E^{(6)}$:
 \begin{equation} e^F=1+F+\frac{1}{2}F\wedge F +\frac{1}{6}F\wedge F \wedge F.\end{equation}
 Naturally, we expect that T-duality will yield branes of lower dimensions.\\

 Let us consider the simplest case of T-duality in a single (the
sixth) direction.  This will be enough to see the main idea, and the
extension of the result to more general geometrical situations is not
difficult.  With the standard rule of permutation between transverse
scalars and gauge fields,
\begin{equation}\Phi^6\longleftrightarrow A_6,\end{equation}
 we end up with a D5-brane equipped with a connection $A'_\mu$, where
 $\mu$ takes values in the remaining directions $\mu=1,\dots, 5$. That
 is to say, connection and curvature $F':= dA'$ are supported on the
 T-dual worldvolume, and $d$ in the definition of $F'$ only
 comprehends the tangent directions to the T-dual world-volume,
 $d=\sum_{\mu=1}^5 dx^\mu \frac{\partial}{\partial x^\mu}$.\\

 However, if we follow these rules to transform the Chern character,
 the result is not the Chern character $e^{F'}$. Of course
 curvatures are wedged with differentials of the transverse scalars,
 thus making odd forms enter the expression, but the zero-order term
 remains untouched:
\begin{equation}e^F= 1+F+\frac{1}{2}F^2+\frac{1}{3!}F^3\end{equation}
\begin{equation}{}\longleftrightarrow 1+d\Phi^6+F'+d\Phi^6\wedge F'+\frac{1}{2} {F'}^2+\frac{1}{2}d\Phi^6\wedge {F'}^2+ \frac{1}{3!}{F'}^3.\end{equation}
The Chern character is therefore not covariant with respect to
T-duality of the six-dimensional object
\begin{equation}(\Phi^a, A_\mu)\end{equation}
describing transverse scalars and gauge fields on the branes.\\

Consider instead, as in (\ref{forms}) the wedge product of the Chern character with the
determinant of the conormal bundle of the brane. Of course this
modification is immaterial in the case of the D6-brane, since the
annihilator has dimension zero, and the corresponding
factor is a differential form with degree zero\footnote{which we
normalize to one; this choice of normalization corresponds to the fact
that the maximal isotropics are associated with pure spinor lines.}:
 \begin{equation}e^F|_{E^{(6)}}=\det(\Ann E^{(6)})\wedge e^F.\end{equation}
But the unit factor is not inert under T-duality.  The expansion of
the T-dual of $e^F$ is now weighted by the T-dual of the determinant
of the annihilator of the six-dimensional space, which is the
one-form $d\Phi^6$. In particular, the zero-order term coming from the
expansion of $e^{F'}$ or from $(e^F)'$ is weighted by a form whose
kernel automatically encodes the location of the brane. Higher-order
terms from the expansion of $(e^F)'$ either contain factors of
$d\Phi^6$ (and therefore do not contribute to an overall wedge product
with $d\Phi^6$), or only contain the reduced field strength $F'$, thus
reproducing the expansion of $e^{F'}$:
\begin{equation}\label{T-duality}\left(\det(\Ann E^{(6)})\wedge e^F\right)'=d\Phi^6\wedge (e^F)'= d\Phi^6\wedge e^{F'}=\left(\det(\Ann E^{(6)})\right)'\wedge e^{F'}.\end{equation}
 This object is the T-dual of $L(E^{(6)},F)$ by the correspondence
 (\ref{forms}), and it is still a pure spinor, precisely the one that
 we would associate to the T-dual of the D6-brane.\\

 We have therefore shown from ordinary Buscher rules that the product
of annihilator and Chern character transforms as D-branes under 
T-duality\footnote{When the brane is non-Abelian or when more than one
T-duality is performed, the formula should be properly
covariantized. The exponentiated field strength should be modified to
include the commutators of the scalars, and the connection on the
normal bundle needs to be included.  We will concentrate mostly on
the simplest case of Abelian D-branes, and the importance of inclusion of
$\det(\Ann E^{(6)})$ should be already clear in this case.}.  This
object makes use all the six-dimensional data $(\Phi^a, A_\mu)$ in any
of its T-dual descendants, whereas the Chern character only made use
of the gauge part.  Thus, starting with a maximally isotropic space
for the $O(6,6)$ pairing (\ref{pairing}) of type zero, corresponding
to the whole six-dimensional space equipped with a two-form, T-duality
acts within maximally isotropic subspaces. Thus, T-duals of D6-branes
are in one-to-one correspondence with pure spinors, the parity of the
type of which is dictated by the number of T-dualities.\\

  The important
point is that the transverse displacements enter symmetrically with
the gauge fields on the brane and as already mentioned the dimension
of the generalized brane is always six \cite{hm}. What changes in
passing from D$p$-brane to D$p'$ is the split between the longitudinal
and transverse directions, or windings and momenta along the worldvolume,  
as encoded in the changes of the type of the
corresponding pure spinors.

\subsection{Stability conditions and pure spinors}

 The properties of generalized submanifolds worked out in
 \cite{gualtieri} correspond to equations of motion (as D-branes can
 be considered as instantons), regardless of the conditions for
 existence of solutions of these equations. Stability equations imply
 the equations of motion and replace their analytical solving by a
 topological problem \cite{UY}.\\
 
For example, let $F$ be the  curvature of a holomorphic line bundle,
  \begin{equation}\label{holomorphicity} F^{(0,2)}=0,
\end{equation}
deriving from an antihermitian connection. In complex dimension $n$,
the self-duality equation of the curvature reads \cite{Atiyah} as the
Hermitian Yang--Mills equation
\begin{equation}\label{HYM}
F\wedge\frac{\omega^{n-1}}{(n-1)!}=c\frac{\omega^n}{n!}{\mathrm{Id}},\end{equation}
where $c$ is a constant proportionality factor.  It turned out in
\cite{MMMS} that demanding that a D-brane preserves supersymmetry
induces deformations of the stability equation (\ref{HYM}). These
stringy deformations, for a D-brane wrapping a complex $n$-dimensional
submanifold $E$ of a Calabi--Yau manifold, take the form
 \begin{equation}\label{MMMS}\mathrm{Im}(e^{i\theta} e^{\omega+F})|_E=0.
\end{equation}
Taking the limit of small field strength gives back (\ref{HYM}) with the identification $c=-i\tan\theta$.\\

The constant phase $\theta$ was shown in
\cite{lyz} to be obtained by mirror symmetry from the constant phase
contained in the special Lagrangian condition for an A-brane wrapping
a submanifold $L$:
 \begin{equation}\label{lagrangian} \omega|_L=0,
\end{equation}
 \begin{equation}\label{special} {{\mathrm{Im}}}(e^{i\theta} \Omega|_L)=0.
\end{equation}
Stability conditions should be rephrased in terms of the generalized
 tangent bundle. Indeed, the stability conditions for D-branes of the
 A and B models \cite{MMMS,kapustin--li} can be encoded in terms of
 pure spinors, and the conditions are exchanged exchanged by mirror
 symmetry \cite{gm}. This exchange incorporates non-Lagrangian
 A-branes into mirror symmetry, and includes the mirror transformation
 that is inverse to the one considered in \cite{lyz}. We will come
 back to this point in section 4, after generalizing the notion of
 stable triples. Stable triples consist of holomorphic line bundles
 $E_1$ and $E_2$, together with a map $T$ between them. As spaces of
 one-forms are fibered over bundles in GCG by the definition
  (\ref{gtbundle}) 
 of generalized tangent bundles, stability conditions
 on triples can be generalized by requiring $T$ to be compatible with
 this fibered structure.\\

\section{Generalized stable triples from the B-model}
\subsection{Lower D-brane charges and tachyon condensation}

 D-branes wrapped on generalized complex submanifolds, in the case of
 zero three-form $H$-flux, should admit a description in terms of  elements in K-theory of the spacetime,
 consistently with tachyon condensation. However, we do {\emph{not}}
 have a low-energy effective action for a generalized brane-antibrane
 pair, such as the Abelian Yang--Mills--Higgs model
\begin{equation}\label{YMH}S= \int_E dx\left( \frac{1}{4} \left(F^{(1)}\right)^2+\frac{1}{4} \left(F^{(2)}\right)^2 + \nabla T \nabla T^\ast+\lambda(TT^\ast-\alpha^2)\right).\end{equation}
 We must instead study maps between generalized tangent bundles.  These maps are the
 only candidates for the tachyon fields in generalized complex
 geometry. We shall first stick to the ordinary complex case of the
 B-model, in order to be able to ensure consistency with
 previously-studied brane condensates in the B-model
 \cite{opw}. Mirror symmetry will be used in the next section to
 investigate pairs of topological brane-antibranes of the A-model as
 generalized complex submanifolds subject to tachyon condensation.\\

We will encounter topological constraints of charge conservation, as
 in the ordinary case.  The charges of brane condensates are
 classified by fundamental groups
  \cite{Witten,Horava}. Consider a brane-antibrane pair
 wrapping some $p$-dimensional submanifold. Its decay into the
 closed-string vacuum is inconsistent with charge conservation, as
 soon as the brane and the antibrane carry bundles with different
 first Chern classes. 
 There exists a non-zero net $(p-2)$-brane charge
 \cite{within}. This is an example of a situation where the
 K-theoretic description of D-branes gives an accurate description of
 the charges that are not necessarily captured by homology.\\

 The connection between the net $(p-2)$-brane charge and the tachyon
 field goes as follows. As the condensate has codimension two in the
 original submanifold, the tachyon field has condensed (to the
 non-zero minimum $\alpha$ of its potential) only along two
 dimensions, which means that it stays at the zero value along the
 condensate. The equation $T=0$ now serves as an equation for the
 submanifold wrapped by the condensate, and the tachyon can be seen as
 a map from the transverse space into the gauge group. In the present
 case, the gauge group is $U(1)$ and the transverse space is
 surrounded by a circle. The behavior of the tachyon at transverse
 infinity must be encoded in a map from this circle into $U(1)$. The
 net charge is therefore classified by $\pi_1(U(1))$. The non-Abelian
 cases correspond \cite{ABS} to higher codimensions, say $2k$, together with a
 winding number in $\pi_{2k-1}(U(2^{k-1}))$. We will consider only the
 case of codimension two and Abelian gauge fields.

\subsection{Morphisms between generalized complex branes}
Let us review the analysis of stable triples given in \cite{opw}. 
Let $E_1$ and $E_2$ be two holomorphic line bundles supported on a
two-plane.  They carry connections $A^{(1)}$ and $A^{(2)}$ with field
strengths $F^{(1)}$ and $F^{(2)}$ corresponding to different first
Chern classes. We may assume for definiteness that $E_1$ is trivial,
but it will be appear that the field $T$ only depends on the difference
between the two Chern classes
\begin{equation}\label{classes}[F^{(1)}]-[F^{(2)}]= [dx^1\wedge dx^2].\end{equation}
 The two bundles, considered as a brane and an antibrane, cannot
 condense into the vacuum since this would violate the conservation of
 D$0$-brane charge. The two-dimensional model is
 exactly solvable, in the sense that holomorphicity, together with
 Hermitian conditions, or vortex equations
\begin{equation}\bar{\partial}\label{holotachyon} T +TA^{(2)}-A^{(1)}T=0,
\end{equation}
\begin{equation}\label{vortex1} ig^{\mu\bar{\nu}} F^{(1)}_{\mu\bar{\nu}}+ TT^\ast\sim {\mathrm{Id}}^{(1)},\end{equation}
\begin{equation}\label{vortex2} ig^{\mu\bar{\nu}} F^{(2)}_{\mu\bar{\nu}}- TT^\ast\sim {\mathrm{Id}}^{(2)},\end{equation}
determine the tachyon profile\footnote{The dimension of the identity
matrices on the {\sc{RHS}} of the vortex equations is the rank of the
gauge group.}. Integrating the holomorphicity equation
(\ref{holotachyon}) yields
 \begin{equation} \partial\ln T= A^{(1)}- A^{(2)}.\end{equation}
Due to the unit difference between first Chern
classes (\ref {classes}), one can gauge the phase of $T$ at infinity as
 \begin{equation}T\sim  f(r) e^{i\theta},\end{equation}
where $\theta$ and $r$ are the polar angle and the radial coordinate
on the plane. At some point $p$ however $\theta$ becomes ill-defined,
and $f$ must have a zero at $p$. The position of $p$ is
interpreted as the position of the condensate D$0$-brane, corresponding
to the difference between the two charges in K-theory associated to
the two bundles. This is summarized by the exact sequence
\begin{equation} \label{sequence} 0 \to \mathcal{O}(0) \buildrel T \over\longrightarrow \mathcal{O}(1) \to  \mathcal{O}_p \to 0.\end{equation}

The three equations (\ref{holotachyon}) ,(\ref{vortex1}) and (\ref{vortex2})
therefore play the same role for the Yang--Mills--Higgs model as the
Hermitian Yang--Mills equations (\ref{HYM}) for holomorphic
bundles. They imply the equations of motion derived from the action
functional (\ref{YMH}), hence the name stable triple for
$(E_1,E_2,T)$.\\

 In order to describe condensates in generalized geometries, we will
  have to look for the zeroes of (generalized) tachyon fields. Before
 investigating maps between generalized complex branes, let us recall
 a formal relationship between transverse scalars and tachyon fields
 that was noted in \cite{variations}. It is based on the similarity
 between the vortex equations (\ref{vortex1}), (\ref{vortex2}) and the
  one (called Hitchin equation) obtained from dimensional
 reduction of Hermitian Yang--Mills equation for non-Abelian gauge
 group. In two complex dimensions the Hitchin equation reads
\begin{equation}\label{Hitchin}  F_{1\bar{1}}+ [X,X^\dagger]=c\, \mathrm{Id}.\end{equation}
  The relationship is formal in the sense that $T$ and $X$ are
 different objects for dimensional reasons: the tachyon in the vortex
 equation makes sense for space-filling D-branes, whereas there is no
 transverse scalar (no Hitchin equation) in that case. Moreover, both
 vortex equations and deformed HYM equations (\ref{MMMS}) we are
 studying vortex systems with Abelian gauge fields, whereas non-zero
 commutators in the Hitchin equation are typical of non-Abelian gauge
 theory. However the generalized geometry does make a difference due to
 all branes having the same dimension, allowing us to write down
 tachyon profiles in terms of transverse displacements, or
 characteristic polynomials of some submanifold. transverse
 displacements are taken as arguments by one-forms spanning the
 annihilator space, and the form do not commute, because of the
 exterior algebra, that still exists in the Abelian case.  The
 determinant of the annihilator of a submanifold $E$ has the same
 zeroes as the characteristic polynomial
\begin{equation} p_X(x)=\det(X-x),\end{equation}
where $X^\mu=x^\mu$ are the equations that define the
submanifold. Transformations of transverse scalars under T-duality
induce transformations of annihilators, and by the same token of
tachyon profiles of the form $T=\det(X-x)$. Zeroes of the T-dual of
$T$ are T-dual to the zeroes of $T$. This fact will be used in the
next section when working out mirror images of stable triples. For the
time being we may  note that any pure spinor written in the
form (\ref{forms}) can be regarded as the result of condensation involving
a tachyon profile whose zeroes are on $E$. What remains to be checked
is that the graph condition defining the generalized tangent bundle is
compatible with such maps.\\

To this end, let us consider the embedding of the two-dimensional model of \cite{opw} in GCG.  
The bundles $TM$ and $T^\ast M$ are therefore separately stable under the action of
 the diagonal generalized complex structure $\mathcal{J}$ 
 \begin{equation}\label{GCstructure}\mathcal{J}=\begin{pmatrix}
J  & 0 \\
0 & -J^\ast \hfill 
\end{pmatrix}\end{equation}
 that comes from an ordinary complex structure $J$ on $M$. Consider a
 GC D2-brane $\mathcal{E}_1$ and a GC anti-D2 brane $\mathcal{E}_2$.
 As reviewed above, the generalized complex branes $\mathcal{E}_1$ and
 $\mathcal{E}_2$ are locally graphs over holomorphic $U(1)$-bundles
 $E_1$ and $E_2$ sharing a base denoted by $N$, which is a submanifold of $M$. The generalized tangent bundles are respectively
\begin{equation}\tau_{N}^{F^{(1)}}=\{X+\xi\in TB_1 \oplus T^\ast M|_{N},\, \xi=\iota_X F^{(1)}\},\end{equation} 
\begin{equation} \tau_{N}^{F^{(2)}}=\{X+\xi\in TB_2 \oplus T^\ast M|_{N},\, \xi=\iota_X F^{(2)}\}.\end{equation} 
Let $T$ be a morphism between the two generalized tangent bundles
 ${\mathcal{E}}_1$ and ${\mathcal{E}}_2$. As we are considering
 Abelian gauge fields, it has to act on the $U(1)$ fibers of $E_1$ and $E_2$ through fiberwise multiplication by a
 scalar function, still denoted by $T$. As usual in the geometry of
 principal bundles, we consider a change from local chart $U_\alpha$
 to local chart $U_\beta$, with the data consisting of the local
 multiplicative function $T$, together with the local gauge potentials
 on the intersection of the two charts. Local gauge potentials are
 pull-backs of the connection one-forms $a^{(1),(2)}$ by local
 sections of the bundles:
\begin{equation}A^{(1)}=s_1^\ast a^{(1)}, \;\;\;\;\;\;\;\;
A^{(2)}=s_2^\ast a^{(2)}.\end{equation} 
Let $g_1$ and $g_2$ be
transition functions of ${\mathcal{E}}_1$ and ${\mathcal{E}}_2$
respectively between the two local charts $U_\alpha$ and $U_\beta$. As
multiplication by $T$ maps the $U(1)$-fiber of ${E}_1$ to the one of
${E}_2$ above the same point $x$, we can obtain a transition function
relating $A^{(2)}$ to $A^{(1)}$, because sections of the two bundles
are exchanged as follows:
\begin{equation}s_2(x)=s_1(x) g_1(x)T(x)g_2(x), \end{equation}
\begin{equation}A^{(2)}=g_2^{-1}T^{-1}g_1^{-1}A^{(1)} g_1 T g_2 + g_2^{-1}T^{-1}g_1^{-1}{\bar{\partial}}(g_1 T g_2),\end{equation}
so that there is a holomorphicity condition on $T$ since the gauge
potentials have definite type, say $(1,0)$:
\begin{equation} {\bar{\partial}}T= TA^{(1)}-A^{(2)}T.\end{equation}
The curvatures $F^{(1),(2)}$ transform in the usual fashion, by
 adjoint $U(1)$ actions performed on the two bundles separately. The
 two Abelian field strengths are therefore invariant under change of
 charts, and so are the defining equations of $\tau_{N}^{F^{(1)}}$ and
 $\tau_{N}^{F^{(2)}}$:
 \begin{equation}\label{fiber} \xi=\iota_X \left(g_i^{-1}F^{(i)} g_i\right)=\iota_X F^{(i)}, \;\;\;\;\;i=1,2.\end{equation}
  Once supplemented with Hermitian conditions such as
 (\ref{vortex1}), (\ref{vortex2}), the holomorphicity condition for the tachyon
 gives again the exactly solvable profile of \cite{opw}. These
 Hermitian conditions, with $T$ the characteristic polynomial of some
 submanifold, make sense in GCG as reduction
 of Hermitian Yang--Mills equations on that submanifold.\\

 The generalization of the stable triple analyzed in \cite{opw}
 induces a map $T$ acting fiberwise on the $U(1)$-bundles carried by
 $E$, such that the following sequence is exact:
\begin{equation} \label{gcsequence} 0 \to \tau_N^{0}  \buildrel T \over\longrightarrow \tau_N^{C} \to \tau_{T^{-1}(0)}^{0}\to 0,\end{equation}
  where $C$ is the difference between the first Chern classes of the
  bundles $E_1$ and $E_2$. Consistently with K-theory, we could add
  the same $U(1)$-bundle with curvature $F$ to $E_1$ and $E_2$, and
  would still have a sequence, with the same $T$, but involving
  $\tau_E^{F}$ and $\tau_E^{C+F}$.  The sequence (\ref{gcsequence})
  embeds stable triples of the topological B-model into generalized
  complex geometry.\\

 Given the earlier discussed T-duality rules, we know that this
 low-dimensional situation can be mapped to brane-antibrane pairs of
 higher (even) dimensions by applying even numbers of T-dualities,
 provided the product of tachyon condensation still has codimension
 two, corresponding to Abelian gauge fields. In particular,
 generalized D4 brane-antibrane pairs with different first Chern
 classes carry a generalized D2-brane charge.\\

\section{More generalized brane-antibrane pairs}

\subsection{Special Lagrangian branes as tachyon condensates}
 Since the generalized complex structure (\ref{GCstructure}) we have
  considered so far is diagonal (i.e. corresponds to the ordinary
  complex structure), we have just given a generalized description of
  a situation already known in the topological B-model. As already
  mentioned, taking an anti-diagonal GC structure should give
  branes of the A-model \cite{gualtieri}. Here we describe the latter
  by applying T-duality, and ask whether we recover expected features
  of stable A-branes. As we shall see, different dimensions of
  A-branes, corresponding to different terms in the expansion of $e^F$
  in the pure spinor involved in a stability condition, are tied
  together in the mirror images of particular stable triples.\\

 The open-string version of the mirror correspondence between pure
 spinors $e^{i\omega}$ and $\Omega$ was derived in~\cite{gm} as:
\begin{equation}\label{mirrorstab}e^{i\omega+F}\longleftrightarrow\Omega \wedge e^{F'}.\end{equation}
 Picking the term of the right degree in the expansion of the
 {\sc{RHS}} in powers of the field strength, one finds either the
 special Lagrangian condition
\begin{equation}{\mathrm{Im}}(e^{i\theta}\Omega|_L )=0,\end{equation}
or the stability condition with non-trivial field strength that occurs
in the case of a stable five-dimensional non-Lagrangian brane $Y$
\cite{kapustin--li}:
\begin{equation}\label{nLag}{\mathrm{Im}}(e^{i\theta}\Omega|_Y\wedge F)=0.\end{equation}
We are going to see that both terms in the expansion of the {\sc{RHS}}
of (\ref{mirrorstab}) contribute, in the situation where a D4
brane-antibrane pair is mapped by mirror symmetry to a brane-antibrane
pair of non-Lagrangian type.\\

 If one starts with a brane-antibrane pair wrapping a two-dimensional
 submanifold, the mirror bound state will be made of three-dimensional
 branes, as is seen by a Fourier--Mukai
 transformation~\cite{vanenckevort,gm} in a local coordinate patch,
 where $x^\mu$ and $y^\mu$, with $\mu=1,2,3$, are respectively
 coordinates on the base and fibers of a $T^3$ fibration. However,
 starting with a D4 brane-antibrane system carrying a net D2-brane
 charge, mirror symmetry can yield a five-dimensional
 brane-antibrane pair that can condense into a special Lagrangian
 brane of the A-model.\\

 In order to explain this better, let us pass to local complex
 coordinates adapted to a $T^3$-fibration, written as
\begin{equation} z^\mu=x^\mu+iy^\mu.\end{equation}
and consider a field strength of type $(1,1)$,
\begin{equation} F=F_{\mu\nu} dz^\mu d\bar{z}^\nu,\end{equation}
which can be rewritten in terms of symmetric and antisymmetric
matrices $\mathcal{S}_{\mu\nu}$ and $\mathcal{A}_{\mu\nu}$:
\begin{equation} F=   F_{\mu\nu}(dx^\mu\wedge dx^\nu -dy^\mu\wedge dy^\nu)+
 iF_{\mu\nu}(dx^\mu\wedge dy^\nu -dy^\mu\wedge dx^\nu)\end{equation}
 \begin{equation} =: \mathcal{A}_{\mu\nu} \left(dx^\mu\wedge dx^\nu + dy^\mu\wedge dy^\nu\right)+
 i\mathcal{S}_{\mu\nu}(dx^\mu\wedge dy^\nu).\end{equation} It can be
observed that the dimensionality of the mirror A-brane is induced by
the rank of $\mathcal{A}$, due to the two field strengths on the A and
B side being exchanged by Fourier--Mukai transform.  When this rank is
zero, Gaussian integration over the $y^\mu$-coordinates leads to a
delta-function with three-dimensional support weighting the result,
thus leading to a Lagrangian brane.  When this rank is two, only the
kernel of $\mathcal{A}$, say $y^1$ up to a change of coordinates, is
integrated out in the form of a delta-function,
\begin{equation}\int_{T_{y}^3} e^{dy^\mu d\tilde{y}_\mu} e^F= \delta(\tilde{y}_1-\mathcal{S}_{\mu\nu}y^\nu) e^{F'},\end{equation}
 and the mirror A-brane is five-dimensional.\\

A source of concern whenever non-Lagrangian A-branes
\cite{kapustin--orlov,clay,lectures} in Calabi--Yau three-folds are
studied, is the lack of five-cycles to support them: there are no
five-dimensional charges in homology if the ambient Calabi--Yau
three-fold is simply connected. However, $b_1=b_5=0$ is a key point in
the argument of \cite{syz}, ensuring that gauge
bundles carried by D6-branes are trivial and therefore have no
moduli. It is therefore questionable whether the assumption of the
existence of a $T^3$-fibration, used to perform mirror-symmetry
transformations in \cite{gm} to obtain objects encoding the stability
condition of non-Lagrangian A-branes, is eventually consistent with
the very existence of these branes.\\

 A way to escape this puzzling situation is provided by the
 interpretation of D-branes as charges in K-theory \cite{Ktheory,clay}
 rather than homology. Mirror symmetry applied to a D4 brane-antibrane
 pair with different first Chern classes as in (\ref{classes}),
 predicts the following configuration of A-branes : the D2-brane
 condensate maps to a Lagrangian D3-brane, carrying trivial field
 strength. This is associated to a charge in homology. It has to come
 from tachyon condensation between a brane-antibrane pair of higher
 dimension. This pair is the mirror of the pair of stable D4-branes.
 It can consist of two stable A-branes of dimension higher than
 three. These branes are coisotropic and
 carry Chern characters subject to the stability condition
 (\ref{nLag}).  The Lagrangian three-brane $L$ must correspond to the
 locus where the tachyon vanishes.\\

 The T-duality transformation of the transverse scalars, once the
characteristic polynomial of the transverse scalars is identified with
tachyons as suggested in \cite{variations}, dictates the T-duality
transformation rule of the tachyon. In particular, the mirror image of
the locus $T^{-1}(0)$ of the B-model is the set of zeroes of the
characteristic polynomial of the transverse scalars of the Lagrangian
image of the D2-brane condensate.  Since this set has codimension two inside
the brane-antibrane pair, the brane $L$ is still compatible with the
$U(1)$ gauge theory and classification of charges by $\pi_1(U(1))$. If
we call ${F'}^{(1)}$ and ${F'}^{(2)}$ the Fourier--Mukai
transforms of the two field strengths we have been considering in the
B-model, then the mirror image of the sequence (\ref{gcsequence})
\begin{equation} \label{Asequence} 0 \to \tau_Y^{{F'}^{(1)}}  \buildrel T \over\longrightarrow \tau_Y^{{F'}^{(2)}} \to \tau_{L}^{[0]}\to 0.\end{equation}
 Generalized stable triples of the B-model therefore correspond
 by mirror symmetry to pairings between stable non-Lagrangian branes
 in K-theory, further unifying all possible types of A-branes.\\

 In either of the two cases, mapping stability conditions according to
 (\ref{mirrorstab}) shows that, starting with a two-dimensional
 (resp. four-dimensional) stable B-brane, one ends up with a special
 Lagrangian (resp. a stable non-Lagrangian) A-brane. A configuration
 with a stable D2-brane condensate (wrapped on a holomorphic cycle
 $C$) from stable four-dimensional brane and antibrane, with first
 Chern classes differing by $[C]$, is of the type studied in the
 previous sections. This shows that the two mappings between pure
 spinors in the presence of gauge fields, namely T-duality
 (\ref{T-duality}) and mirror-symmetric spinors (\ref{mirrorstab})
 ensure that some stable triples of the B-model are transformed by
 mirror symmetry into a special Lagrangian condensate obtained from a
 non-Lagrangian brane-antibrane pair.\\

\subsection{Generalized D$0$ brane-antibrane pairs on a torus}

 In the generalized set-up, variations of dimension of a brane, either
 by T-duality or tachyon condensation, can be captured by taking into
 account a non-vanishing first Chern class, thus giving rise to a
 generalized submanifold. So far we have disregarded any topological
 charge that would arise from the global geometry of the dual space
 fibered over ordinary bundles. This restriction ensured the equation
 (\ref{fiber}) and the simple lift of the vortex equations to GC
 branes. Moving to the opposite case of generalized D$0$-branes, were
 there can be no net first Chern class for dimensional reasons, we are
 now going to generalize D$0$ brane-antibrane systems on a six-torus,
 taking care of the topological effects coming from winding
 numbers. \\

Winding numbers of strings stretched between D$0$-branes on tori have
been shown in  \cite{taylor}, using a description of D$0$-branes
by matrices, to give rise to non-vanishing commutators that correspond
to a field strength on a T-dual brane. This is a manifestation of
T-duality at the level of D-brane field theory. The covering space
${\bf R}^6$ of a six-torus is paved by such tori, and a D$0$-brane
localized on $T^6$ can be equivalently described by copies of the
brane, one on each torus patch, collectively described by a matrix
whose blocks $X^i_{n_1,\dots, n_6}$ are labeled by space-time
direction $i$ and patch numbers $n_1,\dots, n_6$.\\\

   Even in the Abelian case, when blocks have size one, some
   commutators of matrices have a nonzero value, due to winding
   numbers. Consider a pair of D$0$
   brane-antibranes sitting at the same point $p$ on the six torus,
   but between which a string stretches once along two of
   the circles, corresponding to, say, coordinates $x^1$ and
   $x^2$. That is to say, we are looking at the matrix configuration
\begin{equation}X^i_{n_1,n_2,\dots,n_6}=\delta^{i,1}\prod_{k\neq 1}\delta_{n_k,0}+\delta^{i,2} \prod_{k\neq 2}\delta_{n_k,0}\end{equation}
 corresponding to a single winding sector. There is a term in the
 dimensional reduction of ten-dimensional super Yang--Mills to 0+1
 dimension, proportional to the square of the matrix
\begin{equation}Q^{12}=    2\pi R_1+ 2\pi R_2 ,\end{equation}
that can be encoded in the field strength along the D2-brane obtained
by T-dualizing along the first two cycles,
 \begin{equation} F:= \theta^1\wedge\theta^2.\end{equation}
We single out two coordinates in order to keep within the
codimension-two case upon generalization, as will become clear in what
follows.\\

In a generalized description we would like to describe this
 equivalence as a result of tachyon condensation. This involves trading two
 cycles on the dual torus (two form-coordinates), for two vector
 directions supporting a gauge curvature equal to $F$. The forms
 $\theta^1$ and $\theta^2$ would be traded for gauge fields.  This
 indeed makes sense because 
 the graph conditions, up to a factor of one half that can be absorbed
in the projective definition of the corresponding pure spinor
$\theta^3\wedge\dots\wedge\theta^6\wedge e^F$, read in a coordinate
patch as
\begin{equation}\xi|_{T^{-1}(0)}(X)=(\iota_X F)(X)=X_1\theta^2-X_2\theta^1=: A\end{equation} 
where $A$ is a one-form potential whose field strength is equal to
$F$. Trading two form-directions $\theta^1$ and $\theta^2$ for vector
coordinates is just a modification of the splitting between gauge
fields and transverse scalars. The corresponding phase transition
would come from the tachyon profile
 \begin{equation}\label{profile} T=\det\,\left((\phi^1-p^1)(\phi^2-p^2)\right),\end{equation}
where $p^1$ and $p^2$ are the coordinates of the D0-branes on the
 first two circles of the six-torus.  In order to put these fields
 together, we must look for the generalization of the vortex equations
 to the present case. In particular, tachyons now have to act on the
 dual torus to capture the topological charge associated to the
 winding numbers.\\

 The graph condition in the definition of $\tau_p^{[0]}$, which is
\begin{equation}  \xi|_p=0,\end{equation}
is immaterial because $p$ has dimension zero; $p$ corresponds to the
pure spinor $\theta^1\wedge\dots \wedge\theta^6$.  However, describing
the D$0$-branes as generalized objects, we must not forget that there is
a topological charge on the dual torus. Winding
numbers around the first two directions prevent from writing the
one-forms $\theta^1$ and $\theta^2$ globally as exact forms:
\begin{equation}\label{charge} \int_{T^2}\theta^1\wedge\theta^2=1.\end{equation}
This topological charge is not captured by the local graph equations
of the generalized tangent bundle. This has consequences on
generalized tachyons. In the situation where the topological charge
came from the $U(1)$ fibers, we considered tachyons first as maps into
the $U(1)$ fibers, then checked that the graph condition was intact,
and concluded that no action on the form-coordinates of the GC brane
was induced by the topological charge. Now we are in the extreme
situation where the tachyon can only act on the form-coordinates,
because those are the only coordinates on the GC branes.
 We have a dual torus fibered above the point $p$,
with the topological charge given by (\ref{charge}). Encoding this
charge in the two-form
\begin{equation}\label{two-form} \mathcal{F}:=\theta^1\wedge\theta^2,
\end{equation}
we can consider $\mathcal{F}$  as the field strength deriving from a connection
$\mathcal{A}$ on the dual torus:
\begin{equation}\mathcal{F}=: [D_{\mathcal{A}},D_{\mathcal{A}}].\end{equation}

What made the stable triple a solvable system was the net topological
charge $[F^{(1)}]-[F^{(2)}]$, together with holomorphicity and Hermitian
conditions. We have just encoded the winding number in a cohomology
charge on the dual torus. As for holomorphicity and Hermitian
conditions, we are still considering the diagonal GC structure
(\ref{GCstructure}). All the structures we enjoyed on the original
torus are therefore transported on the dual torus, and we can require
the tachyon field to be holomorphic and to act by multiplication on
the circles of the dual torus in a way that satisfies the vortex
equations written in terms of the winding charge, namely 
\begin{equation}{\bar{\partial}} T= T\mathcal{A} ,\end{equation}
\begin{equation} \mathcal{F} +2 TT^\ast=0,\end{equation}
so that we end up with the same system as in section 3, but written in
the dual torus. The fact that the number of objects of dimensions zero
and two are exchanged does not have to appear in the equations, since
the solution for $T$ only depends on the {\emph{difference}} between
the two vortex equations. The same gauge transformations can now be
used to obtain $T$ as the profile (\ref{profile}) with a single pole
on $p$, the position the D$0$-branes.
 Since $T$ acts on the dual torus, its counterpart in real space as a
 map between the D$0$-branes must be the determinant of the Hodge dual
 of $(\phi^1-p^1)\wedge(\phi^2-p^2)$, which is the annihilator of a
 D2-brane carrying field strength $\mathcal{F}$.\\

Having studied the two extreme cases where the topological charge
 comes entirely from the Chern classes or from a winding sector, we
 can address more general cases where both kinds of charges are
 present. Consider a generalized D2 brane-antibrane system wrapping a
 two-torus inside a six-torus with strings stretching between them
 such that there is a winding number on a dual two-torus, together
 with a difference between the first Chern classes $[F^{(1)}]$ and
 $[F^{(2)}]$. The tachyon acts by
 multiplication both in the $U(1)$ and the dual-torus fibers, and the
 vortex equations split between the tangent and cotangent parts. We
 expect a condensate that corresponds to a generalized D$0$-D4
 system. Note that just as other products of tachyon condensation, this system is supersymmetric and thus stable.

\section{Conclusions}

We have considered two simple systems embedded in GCG, namely stable triples (together with their mirror images) and D0 brane-antibrane pairs, whose condensation (or rather expansion) is dictated by topological charges and is consistent with T-duality.\\

The correspondence between GC branes and pure spinors appears to
reproduce T-duality properties expected from D-branes. Moreover, the definition of GC branes by a graph condition over a
 base which is a $U(1)$-bundle by itself allows for defining tachyons
 as fiberwise multiplication.  T-duality performed in three directions
 can then map holomorphic stable triples (solutions to the vortex plus holomorphicity equations)  to a realization of special  Lagrangian branes as a condensate of a non-Lagrangian brane-antibrane
 pair. This is an example of the unifying power of generalized
 geometries.\\

Since all GC branes are of the same dimension ({\emph{from the   generalized viewpoint}}), the momenta and windings along the brane worldvolume enter in a symmetrical way and a map between the tachyon field and the scalars on the D-brane worldvolume arises rather naturally. This symmetry 
predicts condensation of brane-antibrane pairs into a brane that is higher-dimensional
  ({\emph{from the ordinary viewpoint}}) and carries a charge that is
  predicted with T-duality.  In the case of stable triples, the holomorphicity and the difference between Chern classes were the key elements, whereas here the crucial role is played by the flux associated with winding numbers. Indeed, starting from a D0 brane-antibrane pair with non-trivial windings on a torus, the action of the generalized tachyon yields the system of vortex plus holomorphicity equations on the dual torus, corresponding to an expansion into a D2-brane.  In a similar fashion, a D2 brane-antibrane pair with both Chern class difference and nontrivial windings can produce via generalized tachyon condensation a stable D4-D0 system.  While all this is somewhat  reminiscent of the dielectric effect of \cite{myers}, we note that the branes here are perfectly Abelian.\\

In this note, we have concentrated on the simplest situations in order to illustrate the way GCG incorporates the basic features of D-branes. In particular we have not explored any complications due to the non-Abelian dynamics (thus avoiding both multiple branes and condensation to codimension higher than two), or the $\alpha'$ corrections in B-model corresponding to the deformed HYM (\ref{MMMS}). Also we have not considered the effects of the NS two-form, neither when talking about the K-theory classes classifying D-brane charges  nor when discussing T-duality. Considering the twisted K-theory of \cite{BM} and the incorporation of (generic) $B$-field in the discussion of T-duality presents a big challenge. We feel however that these complications
should enrich but not invalidate the picture developed here. \\

\bigskip
\noindent
{\bf Acknowledgements} 
\medskip

\noindent
We would like to thank S. Hellerman, W. Taylor and A. Tomasiello for interesting discussions.
The work of  RM  was partially supported by
INTAS grant, 03-51-6346, CNRS PICS 2530,
RTN contracts MRTN-CT-2004-005104 and
MRTN-CT-2004-503369 and by a EU Excellence Grant,
MEXT-CT-2003-509661.  RM would like to thank KITP, UCSB for hospitality during the course of this work and acknowledge the partial support from NSF grant  PHY99-0794.


\end{document}